\documentclass[journal,tighten,10pt, onecolumn]{IEEEtran}

\ifCLASSINFOpdf
\else
\fi

\usepackage[utf8]{inputenc}
\usepackage[bookmarks]{hyperref}
\hypersetup{colorlinks=true,citecolor=blue,linkcolor=blue,filecolor=blue,urlcolor=blue}
\usepackage{amssymb}
\usepackage{mathtools}
\usepackage{physics}
\usepackage{fullpage}
\usepackage{tikz}
\usepackage{circuitikz}
\usepackage{caption}
\usepackage{subcaption}

\usepackage{amsmath,amsfonts,amssymb,amsthm,mathtools}

\usepackage{cleveref}

\crefname{thm}{Theorem}{Theorems}
\crefname{prop}{Proposition}{Propositions}
\crefname{lem}{Lemma}{Lemmas}
\crefname{cor}{Corollary}{Corollaries}
\theoremstyle{definition}
\crefname{ref}{Remark}{Remarks}

\usepackage[hang,flushmargin]{footmisc}
\usepackage{tikz}
\usepackage{pgfplots}

\usepackage{xifthen} 

\newcommand{\CeaO}[1][]{
	\ifthenelse{\isempty{#1}}
	{\cC^{(1)}_{\mathrm{ea}} }
	{\cC^{(1)}_{\mathrm{ea},#1}}
}

\usepackage{braket}

\renewcommand\bra[1]{{\langle{#1}|}}
\makeatletter
\renewcommand\ket[1]{%
  \@ifnextchar\bra{\k@t{#1}\!}{\k@t{#1}}%
}
\newcommand\k@t[1]{{|{#1}\rangle}}
\makeatother

\begin{document}
\title{Monotonicity Under Local Operations: Linear Entropic Formulas}  
\author{
    
Mohammad~A.~Alhejji$^{1,2}$
        and~Graeme~Smith$^{1,2,3}$ 
  \thanks{\noindent $^{1}$ JILA, University of Colorado/NIST, 440 UCB, Boulder, CO 80309, USA \newline
    \hspace{1cm} $^{2}$ Department of Physics, University of Colorado, 390 UCB, Boulder, CO 80309, USA \newline
       $^{3}$ Center for Theory of Quantum Matter, University of Colorado, Boulder, Colorado 80309, USA
}}
\maketitle
\thispagestyle{empty} 

\begin{abstract}
\begin{center}
All correlation measures, classical and quantum, must be monotonic under local operations. In this paper, we characterize monotonic formulas that are linear combinations of the von Neumann entropies associated with the quantum state of a physical system that has $n$ parts. We show that these formulas form a polyhedral convex cone, which we call the monotonicity cone, and enumerate its facets. We illustrate its structure and prove that it is equivalent to the cone of monotonic formulas implied by strong subadditivity. We explicitly compute its extremal rays for $\textit{n}  \leq $ 5. We also consider the symmetric monotonicity cone, in which the formulas are required to be invariant under subsystem permutations. We describe this cone fully for all $n$. We also show that these results hold when states and operations are constrained to be classical. 
\end{center}
\end{abstract}

\section{Introduction}
How can we measure correlations between spatially separated parties?  Correlations cannot be generated locally so, at a minimum, any measure of correlation must not increase under local operations. More generally, monotonicity under the action of some relevant set of operations is a typical requirement of any resource measure \cite{Quantum_Resource_Theories_1, Quantum_Resource_Theories_2, Quantum_Resource_Theories_3}. This fact has motivated the study and construction of monotonic formulas, or monotones, in both classical and quantum information theory \cite{Polymatroidal_dependence,Entanglement_Monotones}. For example, entanglement measures have to be monotonic under local operations and classical communication \cite{Squashed_Entanglement, Entanglement_measures}. Entropic monotones---monotones that can be expressed in terms of entropy---are especially useful because of its central role in information theory \cite{Shannon_theory,BSST02, HOW07,CT06}.

The von Neumann entropy, $S(\rho) := -\text{Tr} (\rho \log_{2} \rho)$, quantifies the information stored in a quantum state $\rho$ \cite{BS95}.  The entropies of a tripartite state $\rho_{123}$
satisfy strong subadditivity (SSA) \cite{SSA_proof}:
\begin{equation}
S(\rho_{13}) +  S(\rho_{23}) - S(\rho_{3}) - S(\rho_{123}) \geq 0 ,
\end{equation}  
where  $\rho_{13}$:= $\text{Tr}_{2} (\rho_{123})$, $\rho_{12}$:= $\text{Tr}_{3} (\rho_{123})$ and $\rho_{3}$:= $\text{Tr}_{12} (\rho_{123})$ are marginals of $\rho_{123}$.  Strong subadditivity is a fundamental tool in quantum information theory and beyond (cf \cite{BSST02, HOW07, BKN00}). Remarkably, it gives us all known linear inequalities limiting the von Neumann entropy. This raises two questions: \textbf{(i)} What linear entropic monotones does SSA imply? \textbf{(ii)} Do linear entropic monotones exist which are not implied by SSA? We answer these two questions below. For the first question, we provide a characterization of monotones implied by strong subadditivity. We show that the answer to the second question is no, i.e., all linear entropic monotones are implied by strong subadditivity.   


In order to talk about correlation measures, we consider density operators defined over a tensor product of multiple Hilbert spaces. Let $N:= \{1$,...,$n\}$ and $J$ be a nonempty subset of $N$. If $\rho_{N}$ is a density operator that describes the state of $n$ systems, then the state of the systems contained in $J$ is given by the partial trace: $ \rho_{J} := \text{Tr}_{N\setminus J} (\rho_{N})$. For a given $\rho_{N}$ and each nonempty $J \subseteq
N$ we associate an entropy $S(J):=S(\rho_{J})$. We call the tuple $(S(J))_{J \subseteq N}$ the entropy vector of $\rho_{N}$ and think of it as a point in $\mathbf{R}^{2^{n}-1}$. The topological closure of the set of all entropy vectors associated
with $n$-partite quantum states, which we denote by $\overline{\mathbf{A}}_{n}$, is a convex cone \cite{Quantum_Entropy_Inequalities}. That means it is closed under addition and multiplication by non-negative factors. 

We seek formulas $f_{\vec \alpha} $ : $\overline{\mathbf{A}}_{n} \to \mathbf{R}$ which are monotonic under the action of local operations, i.e., local quantum channels, and have the following form:
\begin{equation}
f_{\vec \alpha}(\vec S) := \vec \alpha \cdot \vec S = \sum_{J \subseteq N} \alpha_{J} S({J}). \label{Monotone formula}
\end{equation}
The vectors $\vec \alpha$ also live in $\mathbf{R}^{2^{n}-1}$ and,  for nonempty $J \subseteq N$, we let $\mathbf{M}_{J}$ denote the set of vectors $\vec \alpha$ such that 
$f_{\vec \alpha}$ is monotonic under local processing of the systems in $J$. Henceforth, the word monotone will be used to refer to an element of such sets. 

\begin{figure}[t!]
\hspace*{-1.1cm}  
\includegraphics[scale=0.85, trim = 0cm 18.5cm 3.8cm 4.0cm]{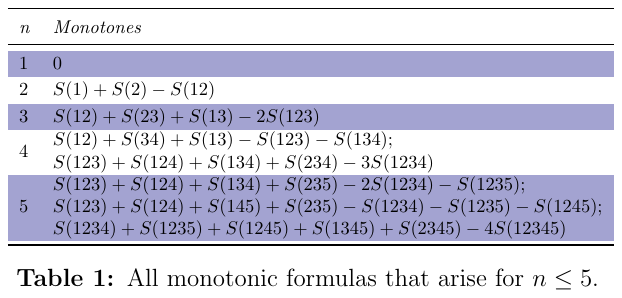}
\captionsetup{labelformat=empty}
\caption{}
\label{fig:table}
\end{figure}

We characterize all monotones by first considering formulas that are monotonic under processing of only one system. There are two fairly simple examples of these:  First, let $J$ and $K$ be
disjoint non-empty subsets of $N$, and let $i \not\in J,K$.  Then strong subadditivity implies that $f_{i,J,K}(\vec{S}) = S(i\cup J)- S(i\cup J\cup K)$ is monotonic under processing of the system labelled by $i$.  Second, any formula that
does not contain entropies involving system $i$ remains the same when only that system is processed. Remarkably, we show below that any formula that is monotonic under processing of $i$ must be a non-negative linear combination of
terms of these two sorts. We can then find the set of monotones under local processing of any subset of subsystems by taking intersections of the appropriate sets for single system processing. This is a rather complicated task,
which we carry out explicitly for up to $n=5$ parties, with the results presented in \ref{fig:table}. For two parties, the mutual information $I(1;2)$ is the unique monotone under local processing, while for larger numbers of
parties we find genuinely new correlation measures.  It remains an open problem to find a general prescription for an exhaustive enumeration of all monotones for an arbitrary number of parties.  

The most important takeaway from this work is that the only monotonic formulas of the form (\ref{Monotone formula}) are the ones implied by strong subadditivity.  It is thought that for $n \geq 4$, there are
linear inequalities that the von Neumann entropy must satisfy which are not implied by strong subadditivity \cite{New_Inequality, robustness_of_QMCs}. Our results indicate that even if this were true,
the corresponding non-negative quantities cannot meaningfully measure correlations. Local operations can cause an increase in whatever resource that these conjectured formulas might quantify. 

The rest of the paper is structured as follows. In Section II, we formalize the posed questions and show that they are equivalent to problems of characterizing convex cones. In Section III, we identify SSA-implied monotones under processing of a single subsystem and prove that a formula is monotonic if and only if its monotonicity is implied by SSA. In Section IV, we study the structure of the monotonicity cone and illustrate its richness. In particular, we provide a table of its extremal rays for $\textit{n}  \leq $ 5. In Section V, we fully describe the symmetric monotonicity cone. That is, the set of monotones which are invariant under subsystem permutations. Section VI concludes with remarks on the consequences of the present results.
\section{Preliminaries}
We now formalize the notion of monotonicity under local operations. Let $\mathcal{N}_{J}$ be a quantum channel, i.e., a linear completely-positive trace-preserving map, that represents an arbitrary processing of a collection of systems in $J \subseteq N$. It is local if it can be written as a tensor product of single system quantum channels, i.e., $\mathcal{N}_{J} = \bigotimes_{j \in J} \mathcal{N}_{j}$. Then a formula $f_{\vec \alpha}$ is monotonic under local processing of $J$ if it satisfies:
\begin{equation}\label{Eq:MONO}
 f_{\vec \alpha}(\vec S(\rho_{N})) \geq  f_{\vec \alpha}(\vec S((\mathcal{N}_{J} \otimes \mathcal{I}_{N\setminus J})(\rho_{N}))) 
 \end{equation}
 for all quantum states $\rho_{N}$ and all local quantum channels $\mathcal{N}_{J}$. Here $\mathcal{I}_{N\setminus J}$ is the identity operation on the systems in ${N \setminus J}$. It immediately follows that the set of monotones under processing of $J$, denoted by $\mathbf{M}_{J}$, is a convex cone in $\mathbf{R}^{2^{n}-1}$. The set of monotones under arbitrary local processing is the convex cone given by,
\begin{equation*}
\mathbf{M}_{N} = \bigcap\limits^n_{i=1} \mathbf{M}_{i},
\end{equation*}
We will characterize $\mathbf{M}_{N}$ by finding the facets and extremal rays of  $\mathbf{M}_{1}$. 

An arbitrary quantum channel $\mathcal{N}$ can be represented as follows:
\begin{equation}
\mathcal{N}(\rho) = \textrm{Tr}_{E}[\mathcal{U}\rho \, \mathcal{U}^{\dagger}], \label{Eq:Isometry}
\end{equation}
where $\mathcal{U}$ is an isometry and E denotes the environment of the channel \cite{Stinespring_dilation}. This leads to the following observation due to Lindblad \cite{lindblad1975}.

\noindent \textbf{Lemma 1}  \textit{A formula $f_{\vec \alpha}$ is monotonic under processing of $1$ if and only if the following inequality holds:}
\begin{equation}
f_{\vec \alpha}(\vec S(\rho_{(1 1') ...n})) \geq f_{\vec \alpha}(\vec S(\rho_{(1) ...n})), \label{Eq: Lindblad}
\end{equation}
\textit{i.e., monotonicity under local operations on $1$ is equivalent to monotonicity under partial trace on $1$.}
\newline
\noindent \textbf{Proof}  To prove necessity, observe that partial trace is itself a local quantum operation. As for sufficiency, consider the representation from (\ref{Eq:Isometry}) and note
\begin{align*}
 f_{\vec \alpha}(\vec S(\rho_{N})) 
&= f_{\vec \alpha}(\vec S((\mathcal{U}_{1} \otimes \mathcal{I}_{N\setminus 1})(\rho_{1 ...n})(\mathcal{U}_{1}^{\dagger} \otimes \mathcal{I}_{N\setminus 1}))) \\ 
&= f_{\vec \alpha}(\vec S(\sigma_{(1E) ...n})) \\ 
&\ge{} f_{\vec \alpha}(\vec S(\sigma_{1 ...n})) \\
&=f_{\vec \alpha}(\vec S((\mathcal{N}_{1} \otimes \mathcal{I}_{N\setminus 1})\rho_{N})),
\end{align*} 
where the first equality is due to the invariance of entropy under the action of isometries.\qed 


\section{The Single System Monotonicity Cone}
We introduce double description (DD) pairs which give a useful description of polyhedral convex cones in real space. A
pair of real matrices $(A, R)$ is called a DD pair if
\begin{equation}
A \vec \alpha \, \geq \, 0  \Leftrightarrow \, \vec \alpha = R \vec \gamma \quad \text{for some} \: \:  \vec \gamma \, \geq \, 0,
\end{equation}
where here $\vec \gamma \geq 0$ means that $\vec \gamma$ has non-negative entries. We say that the rows of $A$ represent the facets of the cone, while the columns of $R$ are its generators. The Minkowski-Weyl theorem states that a cone $\mathbf{C}$ is polyhedral if and only if it is finitely generated \cite{DD_algorithm}. That
is, there exists some real matrix $A$ such that $\mathbf{C}$ $=\{ \vec \alpha \, | \,  A \vec \alpha \, \geq \, 0 \}$ if and only if there exists some real matrix R such that $\mathbf{C}$ $= \{ \vec \alpha \, | \, \vec \alpha = R
\vec \gamma \quad \text{for some} \: \: \vec \gamma \geq \, 0 \}$. If $A$ has full row rank, then a minimal set of generators is unique, up to positive scaling. In that case, $\mathbf{C}$ is said to be a \textit{pointed} cone and there is a one-to-one correspondence between its generators and its extremal rays. Moreover, for each cone $\mathbf{C}$ described by a DD pair $(A,R)$, there is a dual cone $\mathbf{C}^{\vee}$ that is described by the DD pair $(R^{T}, A^{T})$. This fact, which follows from Farkas' lemma, is crucial in proving the present result.

Observe that for any quantum state $\rho_{N}$, processing $1$ in no way affects the entropy of  $K$ if $1 \notin K$. This implies that formulas that have $\alpha_{K} = \pm 1$ for such $K$ and all other entries set to zero span a subset of $\mathbf{M}_{1}$. Additionally, it can be shown via strong subadditivity that for $K \subseteq N$ such that $1 \in K$ and $j \notin K$, the vectors whose only nonzero entries are $\alpha_{K} = - \alpha_{K \cup \{j\}} = 1$ correspond to monotones under processing of $1$. This is another form of the well-known data processing inequality.

Let ${{\mathbf{C}}_1}$ $= \{ \vec \alpha \, | \, \vec \alpha = R_1 \vec \gamma \quad \text{for some} \: \: \vec \gamma \geq \, 0 \}$, where the columns of $R_1$ are the vectors described in the preceding
paragraph. It follows then that ${{\mathbf{C}}_1}$ is contained in $\mathbf{M}_{1}$. As a first step towards showing the opposite containment $\mathbf{M}_{1} \subseteq {{\mathbf{C}}_1}$, we characterize the dual cone $\mathbf{C}_{1}^{\vee}$. Let $\mathcal{P}_{1}(N)$ be the set of all subsets of $N$ that contain $1$ and let it be partially ordered by inclusion. A nonempty family $L  
\subseteq \mathcal{P}_{1}(N)$ is called a lower set of $\mathcal{P}_{1}(N)$ if
\begin{align}
x \in L \Rightarrow  y \in L \; \; \forall  y \subseteq x,
\end{align}
i.e., it is closed under going down in the inclusion order. Upper sets are defined in a similar manner. The complement of a lower set is always an upper set. 
\newline

\noindent \textbf{Theorem 1} \textit{${{\mathbf{C}}_1}$ is the subset of $\mathbf{R}^{2^{n}-1}$  that satisfies}
\begin{equation}
\sum_{J \in L} \alpha_{J} \geq 0   \label{Eq: 3}
\end{equation}
\textit{for all lower sets $L$ of $\mathcal{P}_{1}(N)$ and}
\begin{equation}
\sum_{J \in \mathcal{P}_{1}(N)} \alpha_{J} = 0. \label{Eq: 4}
\end{equation}
\noindent \textbf{Proof} Let ${{\mathbf{C}}_1}' =\{ \vec \alpha \, | \,  A_{1}'\vec \alpha \, \geq \, 0 \}$, where the rows of $A_{1}'$ correspond to the conditions in (\ref{Eq: 3}). We now show that the
extremal rays of ${{\mathbf{C}}_1}'$ are given by the columns of $R_1$ augmented by vectors where the only nonzero entry is $\alpha_{K} = +1$ for $K \in \mathcal{P}_{1}(N)$. Denote this larger generator matrix by $R_1'$. 

Given the fact that ($A_1'$, $R_1'$) is a DD pair if and only if ($R_1'^{T}$, $A_1'^{T}$) is a DD pair, the assertion follows if 
the generators of the cone $\{ \vec \beta \: \: | \: \:  R_{1}'^{T} \vec \beta \: \: \geq \: \: 0 \}$ are the columns of $A_1'^T$. More explicitly, this cone is the set that satisfies
\begin{equation}
 \pm \beta_{K} \geq 0  \text{,} \quad \beta_{J} \geq 0 \quad \text{and} \quad  \beta_{J} \geq \beta_{\{i\} \cup J}
 \end{equation}
 for all $\{i\},J, K \subseteq N$ such that $1\notin K$, $1 \in J$ and $ i \notin J$. The first set of conditions says that the cone is contained in a proper subspace of $\mathbf{R}^{2^{n}-1}$.
 Within this subspace, it is the set that satisfies
 \begin{equation}
 \beta_{J} \geq 0 \quad \text{and} \quad  \beta_{J} \geq \beta_{I} \label{ constraints 6}
 \end{equation}
 for all $J,I \in  \mathcal{P}_{1}(N)$ such that $ J \subseteq I$. We note here that these constraints are nearly identical to the ones satisfied by the quantum relative entropy vector of two states defined over $n-1$ systems. The only difference is the inequality sign is reversed in the second set of constraints in (\ref{ constraints 6}). The extremal rays of the cone of quantum relative entropy vectors, also known as the
 Lindblad-Uhlmann cone, have been explicitly found for all $n$ in \cite{Inequalities_of_relative_entropy}. For self-containment, we reproduce the proof therein. 
 
To enumerate the extremal rays of a pointed polyhedral cone, it suffices to pick subsets of inequalities whose span has codimension 1 and require that they be satisfied with equality. Let $\beta^{*}$ be an extremal ray. Then in addition to satisfying \eqref{ constraints 6}, it is the solution to $2^{n-1} - 1$ linearly independent equations which demand either a component is equal to zero or two components are equal to each other. This means that 
\begin{align}
    \beta^{*}_{J} = 0 \Rightarrow  \beta^{*}_{I} = 0,
\end{align}
for $I,J \in \mathcal{P}_{1} (N)$ such that $J \subseteq I$. Hence, there exists an upper set $U$ such that $\beta^{*}_I = 0$ if and only if $I \in U$. Let $L$ the complement of $U$ in $\mathcal{P}_{1} (N)$. Note that it cannot be empty as $\beta^{*}$ is nonzero by definition. It must be the case that $\beta^{*}_I = \beta^{*}_J = \lambda$ for all $I,J \in L$ and some $\lambda >0$, because of the extremality of $\beta^{*}$ and the fact that it satisfies equations of the form $\beta^{*}_{I} = \beta^{*}_{J}$. For the converse, consider a lower set $L$ that contains $|L|$ subsets and let $\beta^{*}$ be the vector with components equal to 1 for subsets in $L$ and zero otherwise. Observe that $\beta^{*}$ satisfies $2^{n-1} - |L|$ linearly independent equations of the form $\beta^{*}_{I} = 0$ for $I \notin L$ in addition to $|L| - 1$ linearly independent equations of the form $\beta^{*}_{I} = \beta^{*}_{J}$ for $I,J \in L$. 

 Finally, since ${{\mathbf{C}}_1} \subseteq {{\mathbf{C}}_1}' $, then any $\vec \alpha \in {{\mathbf{C}}_1} $ must satisfy the inequalities in (\ref{Eq: 3}). Moreover, $\vec \alpha$ satisfies (\ref{Eq: 4}), as it is a positive combination of only the columns of $R_1$.\qed
\newline

To show that monotones under processing of $1$ must satisfy \eqref{Eq: 4}, consider the classical state of $n$ random variables which are distributed according to the joint probability distribution $ p(x_1,x_2, ... , x_n) = p(x_1) p(x_2, ... ,x_n)$,
where $p(x_2,...,x_n)$ is a deterministic probability distribution. Evaluating an arbitrary linear entropic formula $f_{\vec \alpha}$ on its (Shannon) entropy vector gives $(\sum \alpha_J) H(X_1)$, where the sum is over $J \in \mathcal{P}_{1}(N)$. Since it is always possible to inject more entropy into $1$ via local operations, such a formula is monotonic under processing of $1$ if and only if it is identically-zero. Hence, the desired equality holds.

As for the inequalities in \eqref{Eq: 3}, observe that Lemma 1 implies the following. To show the inequality corresponding to a lower set $L$ is satisfied by all elements in $\mathbf{M}_{1}$, it suffices to find states with entropy vectors that satisfy
\begin{align}
    S(Q | J) = c \quad \text{and} \quad S(Q | K) = 0, \label{secret sharing}
\end{align}
for some positive constant $c$ and for all $J \in L$ and  $K \in U = \mathcal{P}_{1} (N) \setminus L$. Here, $Q$ is a stand-in for the part of $1$ to be discarded in some processing. Note that system $1$ goes for the ride in these constraints and so we may assume that it is independent of all else. From now on, we invoke the isomorphism $\mathcal{P}_{1} (N) \cong 2^{N \setminus 1}$.  

To demonstrate a state that realizes such an entropy vector, we briefly expose the theory of classical secret-sharing schemes \cite{How_to_Share, Blakley1979}. In a secret-sharing scheme, there is a dealer who wishes to distribute shares of a secret $Q$, which may be modeled as a discrete finite random variable, among a party of $n$ individuals such that two conditions are met: \textbf{(i)} (correctness) if a subset of individuals is authorized, then by pooling their shares together, they can recover the secret faithfully. \textbf{(ii)} (perfect privacy) if a subset is not authorized, then the individuals in it cannot learn anything about the secret from their shares. The family of authorized subsets in a secret sharing scheme is called the access structure of the scheme. Access structures are naturally required to be upper sets of $2^{N}$ under the inclusion order. Secret sharing schemes with arbitrary access structures were first explicitly constructed by Ito, Saito and Nishizeki in \cite{Secret_Sharing}. Following their construction, let $Q$ be a uniformly distributed binary random variable and let $L$ be a given lower set of $2^{N \setminus 1}$. For each $K = \{k_{1}, ..., k_{s}\}$ in $U = 2^{N \setminus 1} \setminus L$, we do the following:  \textbf{(i)} choose $s-1$ bits independently and uniformly randomly $b_{1}, ..., b_{s-1}$. \textbf{(ii)} let $b_{s} = q \oplus b_{1} \oplus ... \oplus b_{s-1}$, where $\oplus$ denotes addition mod 2. \textbf{(iii)} give the bit $b_{i}$ to the individual $k_{i}$. If $s = 1$, then $K$ is an authorized singleton and so we let individual $k_{1}$ have the secret. Depending on $U$, it could happen that certain individuals have more shares than others. Note that this process could be made more efficient by restricting it to minimal elements of $U$ under the inclusion order.
In any case, we clearly have $S(Q|K) = 0$ for all $K \in U$. On the other hand, for $J \notin U$, there is always at least one ``missing'' bit in all the available shares and so $S(Q|J) = 1$. This classical state of $n+1$ random variables realizes the desired entropy vector \eqref{secret sharing}. Below, we show explicit constructions of these states for two different access structures. For more details about secret-sharing schemes, see the recent survey \cite{Secret_Survey}.

\begin{figure}[b!]
\centering
\begin{subfigure}{.5\textwidth}
  \centering
  \includegraphics[width=.75\linewidth]{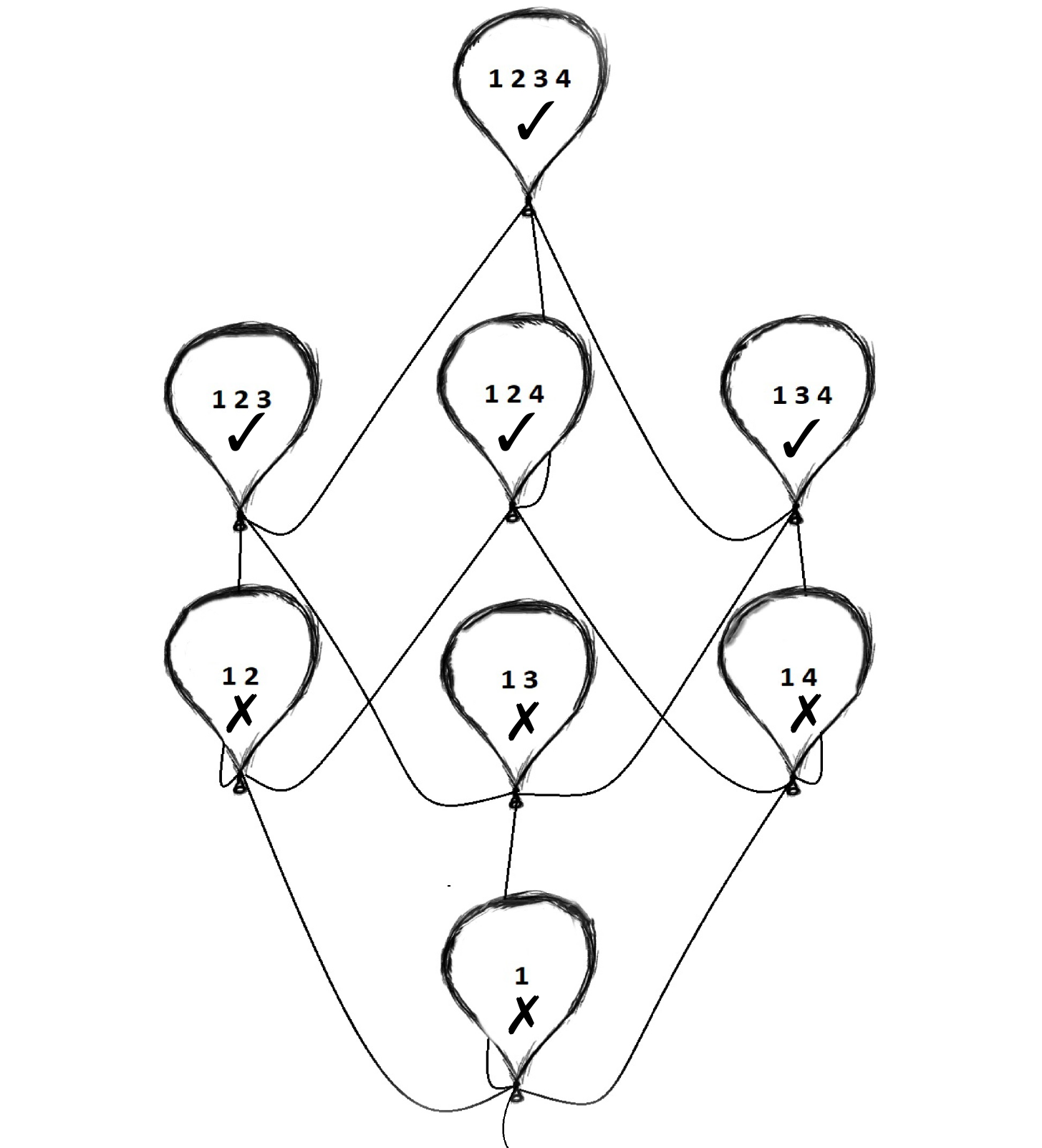}
  \caption{}
  \label{fig:sub1}
\end{subfigure}%
\begin{subfigure}{.5\textwidth}
  \centering
  \includegraphics[width=.75\linewidth]{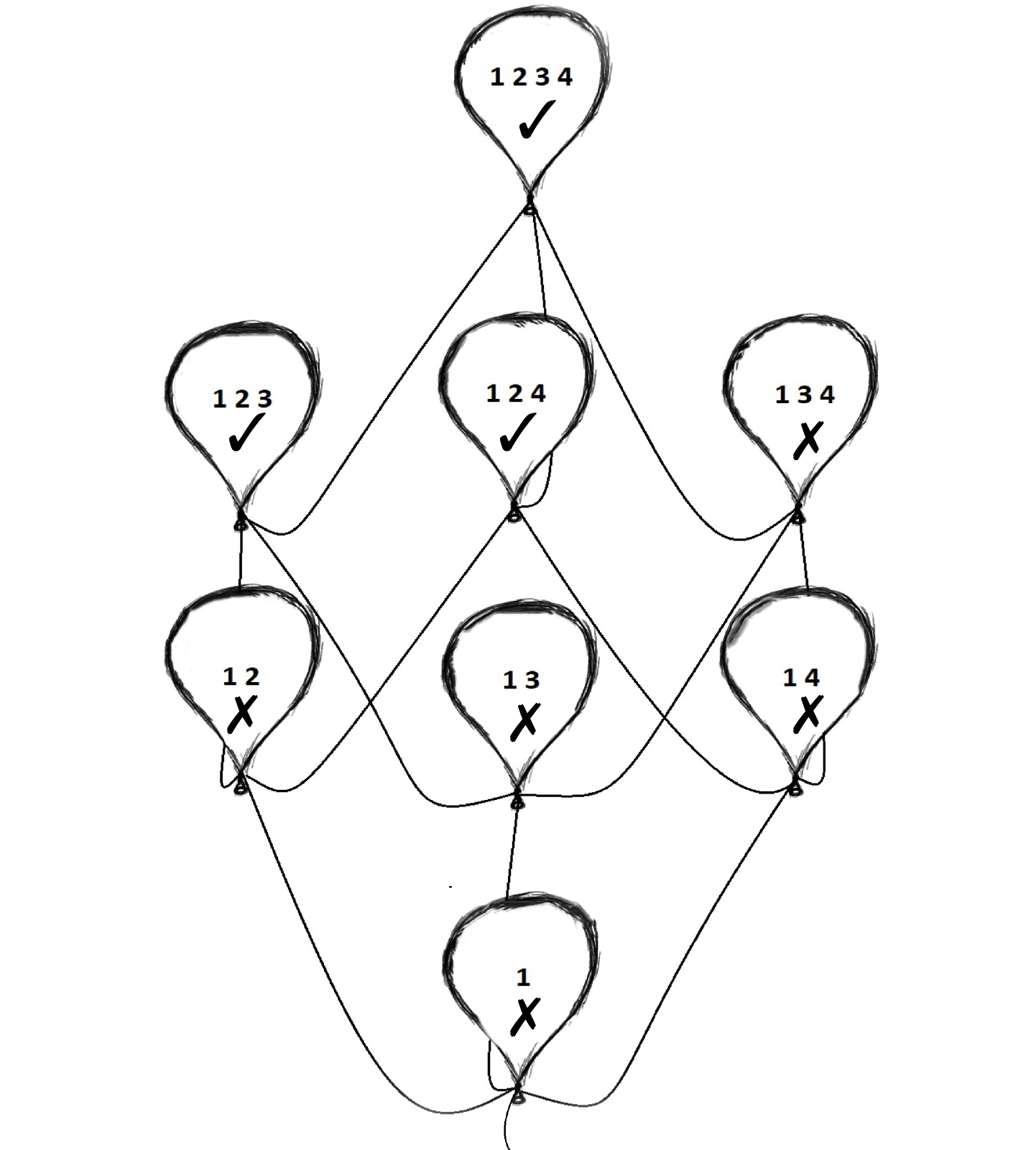}
  \caption{}
  \label{fig:sub2}
\end{subfigure}
\caption{\textbf{(a)} A threshold access structure which corresponds to the inequality $\alpha_{1} + \alpha_{12} + \alpha_{13} + \alpha_{14} \geq 0$. \textbf{(b)} A more complex access structure which corresponds to the inequality $\alpha_{1} + \alpha_{12} + \alpha_{13} + \alpha_{14} + \alpha_{134} \geq 0$.}
\label{fig:test}
\end{figure}
\newpage
\noindent \textbf{Example 1} 
The access structure in \ref{fig:sub1} is called a threshold access structure. That is, if the number of individuals in a given collection exceeds a certain number, which in this case is 2, then they are able to recover the secret. Otherwise, they cannot learn anything about it. A state that realizes such a scheme is:
\begin{align}
    \rho^{a}_{Q234} = \frac{1}{16}\sum \; \ket{i} \bra{i}_{Q} \otimes \ket{j,k} \bra{j,k}_{2} \otimes \ket{j\oplus i,l} \bra{j\oplus i,l}_{3} \otimes \ket{k\oplus i, i \oplus l} \bra{k\oplus i, i \oplus l}_{4},
\end{align}
where the sum is over $i,j,k,l \in \{0,1\}$. Clearly, any two individuals can recover $Q$ exactly, while any one individual cannot.\newline
\noindent \textbf{Example 2} 
The access structure in \ref{fig:sub2} is more hierarchical. Individuals 3 and 4 cannot recover the secret without the help of individual 2, but the latter needs only one of them to recover the secret. We can realize this secret-sharing scheme by the following state:
\begin{align}
    \rho^{b}_{Q234} = \frac{1}{16}\sum \; \ket{i} \bra{i}_{Q} \otimes \ket{j,k} \bra{j,k}_{2} \otimes \ket{j\oplus i} \bra{j \oplus i}_{3} \otimes \ket{k \oplus i} \bra{k \oplus i}_{4},
\end{align}
where again the sum is over $i,j,k,l \in \{0,1\}$. 

\section{The Monotonicity Cone}
The results of the preceding section imply that $\vec \alpha \in \mathbf{M}_{1}$ if and only if $f_{\vec \alpha}$ admits the following representation:
\begin{equation*}
f_{\vec \alpha} (\vec S) = -\sum_{j\in N, I \subseteq N} v_{j,I}  S(j | 1 \cup I)  + \sum_{I \subseteq N} w_{I} S(I),
\end{equation*}
where $j \ne 1$, $1 \notin I$ and  $v_{j, I} \geq 0$. Therefore, the monotonicity of $f_{\vec \alpha}$ under local operations is equivalent to the existence of such a representation for all $n$ subsystems, which in turn is equivalent to $\vec \alpha$ simultaneously satisfying the conditions mentioned in Theorem 1 for all $n$ subsystems. In particular, this says that all monotones must be \textit{balanced}. A formula is balanced if it satisfies all versions of Eq.~(\ref{Eq: 4}). That is, the sum of all components $\alpha_I$ such that $i \in I$ must vanish for all $i \in N$. 

For $n=1$, no monotones exist as any mixed quantum state can be processed into having a higher or lower entropy. 

 As for $n=2$, only one balanced formula exists, up to positive scaling, and it is
the mutual information. 
\begin{equation*}
I(1;2):= S(1) + S(2) - S(12).
\end{equation*}
It obviously satisfies the inequalities associated with processing on 1, likewise for 2, and so is indeed a monotone. This can also be seen as a direct consequence of SSA which asserts the non-negativity of the quantum conditional mutual information $I(1;2|3):= S(13) + S(23) -S(3) -S(123)$.

The case of three systems is more interesting. The following monotone appears:
\begin{equation*}
J(1;2;3):= S(12) + S(23) + S(13) - 2S(123).
\end{equation*}
Observe that it vanishes if and only if the tripartite state is a product state, which indicates that it measures some genuine symmetric three-way correlations. It is in fact the quantum mechanical version
of Han's \textit{dual total correlation} for three random variables \cite{Dual_Correlation}. An operational interpretation of this quantity remains elusive both in the classical and quantum settings. However, it has been used to obtain bounds on distillation rates in certain classical and quantum cryptographic schemes \cite{Secrecy_Monotones}. 

The first novel monotone arises in the case of four systems:
\begin{equation*}
U(1;2;3;4):= S(12) + S(34) + S(13) - S(123) - S(134).
\end{equation*}
It is not immediately obvious what to make of this asymmetric quantity, but seeing that it is equal to both $I(2;3|1) + I(1;34)$ and $I(1;4|3) + I(3;12)$, we suspect that it measures some kind of four-way
correlation along the 12$|$34 partition. We note that enumerating the extremal rays of $\mathbf{M}_{N}$ for large $n$ seems to be a highly non-trivial task and leave it as an open problem. Below is a table of all monotones, up to system
permutations, for $n\leq5$.


\section{The Symmetric Monotonicity Cone}
The problem of finding entropic monotones can be made considerably simpler by requiring invariance under single-system permutations. This is equivalent to imposing the following set of conditions on $\vec \alpha$:
\begin{equation*}
\alpha_{I} = \alpha_{I'} = a_{i}
\end{equation*}
for all $I, I' \subseteq N$ that have the same number of elements $i$. For a given number of subsystems $n$, monotonic formulas that satisfy these conditions form a polyhedral convex cone that is properly contained in a subspace of dimension $n$. Moreover, its facets are far fewer than the ones of the monotonicity cone. \\

\noindent \textbf{Lemma 2} \textit{The symmetric monotonicity cone is the set in $\mathbf{R}^{n}$ that satisfies:}
\begin{align*}
a_{1} + \binom{n-1}{1} a_2 + ... + \binom{n-1}{k-1} a_k \geq& \:0, \\
a_{1} + \binom{n-1}{1} a_2 + ... + \binom{n-1}{n-1} a_n = &  \:0, 
\end{align*}
\textit{where $ 1 \leq k \leq n-1$.}\\
\noindent \textbf{Proof} We remark that the coefficient multiplying $a_i$ is the number of subsets of a set of $n-1$ elements which contain $i-1$ elements, i.e., $(i-1)$-sets.  Recall that we use the isomorphism $\mathcal{P}_{1} (N) \cong 2^{N \setminus 1}$. Once symmetry is imposed, observe all versions of the equality (\ref{Eq: 4}) boil down to the equality above. Next, note that the inequalities above are independent and implied by monotonicity plus symmetry. It remains to show that they are satisfied by all symmetric monotones. 

In the $l$th inequality above, denote the quantity on the left-hand side by $A_{l}$. We proceed via induction. That $ a_1 \geq 0$ is immediately evident. Consider the inequalities associated with lower sets of $2^{N \setminus 1}$ which have subsets which contain at most one element. Then symmetry implies that $a_1 + a_2 \geq 0$, $a_1 + 2a_2 \geq 0$,..., $a_{1} + (n-1) a_2 \geq 0$ all hold. However, it can be easily seen that the last inequality in conjunction with the non-negativity of $a_{1}$ imply the rest. With this in mind, assume for the inductive step that the first $k$ inequalities above imply all inequalities associated with lower sets which contain at most subsets of cardinality $k-1$. Given an arbitrary lower set $L$ of $ 2^{N \setminus 1}$ which contains subsets of at most $k$ elements, the associated inequality is
\begin{align}
    a_{1} + \#_{2} a_2 + ... + \#_{k+1} a_{k+1} \geq 0, \label{sym mon}
\end{align}
where $\#_{i}$ denotes the number of subsets of cardinality $i-1$ in $L$. First, we note that if subsets of cardinality $k$ are excluded from $L$, we get another lower set which contains subsets of at most $k-1$ elements. By the inductive hypothesis, the associated inequality can be written as follows:
\begin{align}
 a_{1} + \#_{2} a_2 + ... + \#_{k} a_{k} = \sum_{i=1}^{k} \gamma_{i} A_{i} \geq 0,     
\end{align}
where $\gamma_{i} \geq 0$ and $\gamma_{k} = \frac{\#_{k}}{\binom{n-1}{k-1}}$. We will need an observation due to Sperner \cite{Sperner1928}, 
\begin{align}
    \#_{k} ((n -1) - (k-1)) \geq \#_{k+1} k. \label{sperner}
\end{align}
To see why this inequality holds, observe that each $k$-set contains $k$ $(k-1)$-sets and so $\#_{k+1} k$ is the number of $(k-1)$-set instances in the $k$-sets within $L$, including possible duplicates. Since $L$ is a lower set, all those $(k-1)$-sets are also in $L$. Since each $(k-1)$-set is contained in $((n-1)-(k-1))$ $k$-sets, then that number of instances is bounded from above by $\#_{k} ((n -1) - (k-1))$. \newline
If we let
\begin{align}
 \quad \eta_{i} = \gamma_{i} \quad \text{for} \quad 1 \leq i \leq (k -1), \quad 
    \eta_{k} = \frac{\#_{k}}{\binom{n-1}{k-1}} - \frac{\#_{k+1}}{\binom{n-1}{k}} \quad \text{and} \quad  \eta_{k+1} = \frac{\#_{k+1}}{\binom{n-1}{k}},
\end{align}
then we have $\eta_{i} \geq 0$ for all $i$, where we used \eqref{sperner} to show $\eta_{k} \geq 0$. Furthermore, we have 
\begin{align}
    \sum_{i=1}^{k+1} \eta_{i} A_{i} = a_{1} + \#_{2} a_2 + ... + \#_{k+1} a_{k+1}.
\end{align}
Hence, the assertion follows by induction. \qed
\\



Therefore, the symmetric monotonicity cone is defined by one equality and $n-1$ inequalities, which is significantly less complex than the monotonicity cone. So much so that we can solve for its extremal rays
for arbitrary $n$. \\ 

\noindent \textbf{Theorem 2} \textit{The generators of the symmetric monotonicity cone for $n$ systems are unique (up to positive scaling) and can be spanned by $n-1$ vectors whose sole nonzero elements are:}
\begin{align}
a_l = \frac{1}{l} \quad \text{and} \quad \: a_{l+1} = -\frac{1}{n-l},
\end{align}
\textit{where $1\leq l \leq n-1$. Written in terms of entropies, for each $n \geq 2$, we have the symmetric monotones}
\begin{align}
    (n-l)\sum_{|K| = l}  S(K) - l \sum_{|K| = l+1}  S(K),
\end{align}
where $K \in 2^{N}$.
\\ \par
\noindent \textbf{Proof} To see that the generators are unique, note that a matrix whose rows represent any $n-2$ inequalities of Lemma 2 in addition to the equality therein has rank $n-1$. Consequently, a 1-dimensional subspace, i.e., an extremal ray, is completely specified when only one inequality is allowed to be non-binding. Let it be the $l$th one. Then it is clear that $a_k = 0$ for all $k < l$. Furthermore, $a_l \geq 0$ and given that the ($l+1$)th inequality is binding, we have:
\begin{equation*}
 \binom{n-1}{l-1} a_{l} + \binom{n-1}{l} a_{l+1} = 0
 \end{equation*}
 which implies that $a_k=0$ for all $k > l+1$ as well. Hence, the proposed vectors indeed span the extremal rays of the symmetric monotonicity cone. \qed 

\section{Concluding Remarks}
We have systematically studied the cone of multipartite linear entropic formulas that are monotonic under the action of local quantum channels. For two subsystems, the mutual information is the unique linear entropic monotone. For higher numbers of parties, the resulting quantities form a natural family of measures of multipartite correlations. 

One consequence of this characterization is the following observation. An entropic formula is monotonic only if strong subadditivity implies its non-negativity, as each party may choose to erase its own subsystem. That is, any linear entropic inequality which is independent of the non-negativity of conditional mutual information, i.e., a non-Shannon type inequality, cannot correspond to a monotonic formula. As an illustration, consider an instance of the first discovered non-Shannon type inequality proven to hold classically by Zhang and Yeung in 1997 \cite{Non_Shannon},
\begin{align*}
2 I(1;2|3)  + I(1;3|2) + I(2;3|1)
 + I(1;2|4) +  I(3;4) - I(1;2) \geq 0.
\end{align*}
The formula on the left-hand side, while evidently balanced, is not monotonic under local processing by any party. The same can be said about any and all inequalities which are independent of strong subadditivity. It seems it is this independence of strong subadditivity that makes it a challenge to find operational meaning in these non-Shannon inequalities.

\section*{Acknowledgments}
The authors would like to thank Andreas Winter for invaluable discussions on the problem of enumerating the extremal rays of polyhedral convex cones. The authors are also grateful for the advice of the anonymous reviewers which helped correct errors in the manuscript and improve the overall presentation of the material. This work was supported by NSF CAREER award CCF 1652560.

\bibliographystyle{IEEEtran}
\bibliography{IEEEabrv,Monotonicity_2020_draft.bib}
\end{document}